\documentclass[conference]{IEEEtran}
\usepackage{color,graphicx}
\usepackage{hyperref}
\usepackage{float}
\usepackage{amsmath}
\usepackage{caption}
\usepackage{subcaption}
\usepackage{epstopdf}
\usepackage{epsfig}
\definecolor{Gray}{gray}{0.9}
\usepackage[table]{xcolor}
\usepackage{comment}

\usepackage{xcolor}
\usepackage{tikz}

\ifCLASSINFOpdf
\else
\fi

\makeatletter

\newcommand{\Rmnum}[1]{\expandafter\@slowromancap\romannumeral #1@}
\makeatother
\begin{document}
\title{Measurement-based coexistence studies of LAA \& Wi-Fi deployments in Chicago}
\author{Vanlin Sathya, Muhammad Iqbal Rochman, and Monisha Ghosh
\IEEEauthorblockN{}
\IEEEauthorblockA{University of Chicago, Chicago, Illinois-60637
\\
{Email: \{vanlin, muhiqbalcr, monisha\}@uchicago.edu}}
}
\maketitle

\begin{abstract}
LTE-Licensed Assisted Access (LAA) networks are beginning to be deployed widely in major metropolitan areas in the US in the unlicensed 5 GHz bands, which have existing dense deployments of Wi-Fi as well. Various aspects of the coexistence scenarios such deployments give rise to have been considered in a vast body of academic and industry research. However, there is very little data and research on how these coexisting networks will behave in practice. The question of ``fair coexistence'' between Wi-Fi and LAA has moved from a theoretical question to reality. The recent roll-out of LAA deployments provides an opportunity to collect data on the operation of these networks as well as studying coexistence issues on the ground. In this paper we describe the first results of a measurement campaign conducted over many months, using custom apps as well as off-the-shelf tools, in several areas of Chicago where the major carriers have been expanding LAA deployments. The measurements reveal that coexistence between LAA and Wi-Fi in dense, urban environments where both systems aggregate multiple channels, continues to be a challenging problem that requires further research.

\end{abstract}


\section{Introduction}
The deployment of cellular technologies in the unlicensed frequencies has been an active area of research in both academia and industry for a few years now, starting with the proposal for LTE-U from the LTE-U Forum \cite{forum}. Standardization by 3GPP followed and LTE-Licensed Assisted Access (LAA) was specified in Release 13 \cite{3gpp}. The motivation for developing these specifications, which enable cellular technologies to utilize the almost 500 MHz of bandwidth available in the unlicensed 5 GHz band, is the ever increasing throughput requirements of existing networks due to the proliferation of wireless-enabled devices combined with increasing popularity of high-bandwidth mobile applications, such as live streaming and video downloads.\footnote{This work supported by the National Science Foundation (NSF) under Grant No. CNS-1618920.}

The rules for unlicensed device operation in 5 GHz in the US do not mandate the use of a contention-based protocol for medium access, such as the listen-before-talk (LBT) based Carrier Sensing Multiple Access with Collision Avoidance (CSMA/CA) protocol used by Wi-Fi. Hence, the first specification to be developed, LTE-U, did not incorporate any LBT mechanism and instead relied on a duty cycle approach for coexisting with Wi-Fi, which is the dominant, pre-existing user of this band. However, the most commonly deployed specification for LTE in unlicensed bands today is LAA which does implement a LBT protocol similar to that used by Wi-Fi, with different values for parameters such as sensing threshold and transmission intervals. The interplay of these values with those adopted by Wi-Fi determines whether the two systems coexist fairly or not in a given scenario.

\begin{figure*}[t]
  \includegraphics[width=7.1in]{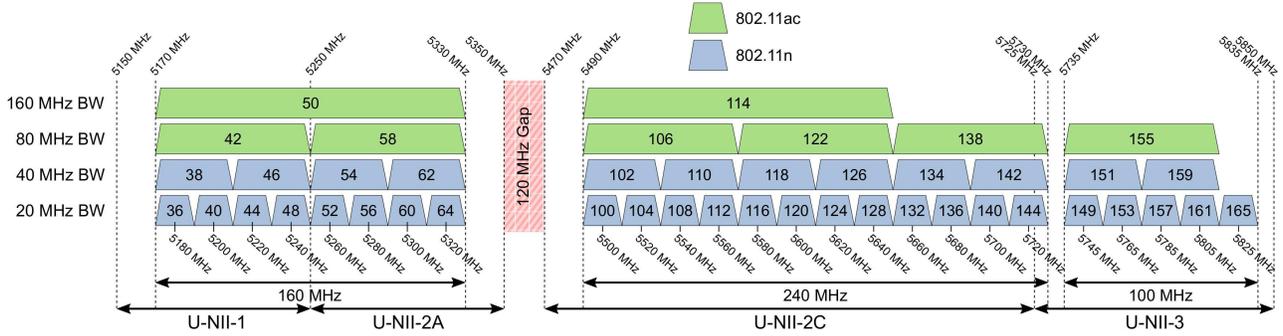}
  \caption{Wi-Fi Channelization in the 5 GHz U-NII Bands}
  \label{fig:chdiag}
\end{figure*}

Most studies of LAA and Wi-Fi coexistence are based primarily on theoretical analyses \cite{TON,tao2015enhanced}, simulations \cite{iqbal2017impact}, and limited experiments \cite{xu2017opportunistic,TCCN}. With widespread deployments of LAA beginning in major cities in the US \cite{ACM}, it is now possible to verify deployment parameters and their impact on coexistence performance. In this paper, we present the first such measurements in various locations in Chicago where LAA networks have been deployed by AT\&T, T-Mobile and Verizon in close proximity to Wi-Fi networks. We believe that this is the first such measurement campaign that studies the effect of LAA deployments on Wi-Fi's performance in practice.

The paper is organized as follows. Section II provides a brief overview of current research, 5 GHz coexistence, focusing on 5 GHz channelization and access methodologies used by LAA and Wi-Fi, and multi-channel operation. Section III describes the data collection approach adopted in this paper including descriptions of LTE \& Wi-Fi tools and data visualization approaches. Section IV provides an overview of LAA deployments observed in the Chicago area followed by Section V which describes in detail the measurements and accompanying discussions at one illustrative location. Finally, conclusions and future research directions are presented in Section VI.

\section{Overview of LAA and Wi-Fi Coexistence}

Coexistence of LAA and Wi-Fi in the unlicensed 5 GHz band has been comprehensively researched and hence we refer the reader to \cite{TON, tao2015enhanced, xu2017opportunistic, zinno2018fair} for details. In this section we present a brief overview of the relevant features pertinent to this paper.

\subsection{Summary of current research on LAA/Wi-Fi coexistence}
In \cite{xu2017opportunistic}, the authors explore the relative issue of differing channelization's (\emph{i.e.,} channel bandwidth asymmetry) between LTE and Wi-Fi, and show that Wi-Fi performance is dependent on the location of the LAA frequency band relative to Wi-Fi’s 20 MHz channel. In~\cite{iqbal2017impact}, Rochman et al. explored the effect of the energy detect (ED) threshold on Wi-Fi and LAA via extensive simulations and demonstrations. If both Wi-Fi and LTE employed a sensing threshold of -82 dBm to detect the other, the overall throughput of both coexisting systems improved, leading to fair coexistence. \cite{7460494} explores design aspects of LBT schemes for LAA as a means of providing equal opportunity channel access in the presence of Wi-Fi. Similarly, \cite{tao2015enhanced} proposed an enhanced LBT algorithm with contention window size adaptation for LAA to achieve fair channel access and Quality of Service (QoS) fairness. In  \cite{cano2015coexistence}, Cano and Leith derived the proportional fair rate allocation for Wi-Fi/LAA (as well as Wi-Fi/LTE-U) coexistence. In \cite{kwan2015fair}, the fairness in the coexistence of Wi-Fi/LAA LBT based on the 3GPP criteria is investigated through a custom-built event-based system simulator.

In our prior work \cite{TON}, we developed an analytical model that predicts the impact of different sensing duration of Wi-Fi and LAA on respective system throughput during coexistence. We validated the model via a National Instrument (NI) test-bed. Further, we compared the fairness performance between the 3GPP definition to a proportional fairness scheme, which guarantees the same fraction-of-time access for all nodes \cite{TCCN}. The results conclusively show that proportional fairness is a much better notion than 3GPP fairness and produces equitable results for both networks on a wider variety of scenarios.

\subsection{5 GHz Channelization}

Fig.~\ref{fig:chdiag} illustrates the channelization used by Wi-Fi in 5 GHz, spanning 5.15 GHz to 5.85 GHz. In the US, these are designated as U-NII bands (Unlicensed National Information Infrastructure) and divided into three categories with different usage rules. U-NII-1 and U-NII-3 bands do not have any restrictions on usage, other than transmit power limitations. However, since radar systems are also deployed in the U-NII-2 bands as primary users, unlicensed devices that wish to use U-NII-2 are required to implement Dynamic Frequency Selection (DFS) whereby the incumbent radar signal has to be sensed and if detected at a certain level the unlicensed device needs to vacate that frequency according to a timing protocol specified by rule. Since these procedures add additional complexity to devices, the U-NII-2 band is sparsely used by Wi-Fi. Thus, even though there is a total of about 500 MHz available, only the U-NII-1 and U-NII-3 bands (a total of 160 MHz) are heavily used. Wi-Fi uses 20, 40, 80, and 160 MHz wide channels, numbered as shown in Fig.~\ref{fig:chdiag}, whereas LAA has been specified only for 20 MHz channels which can be aggregated up to a maximum of three 20 MHz channels.

\subsection{LAA and Wi-Fi Medium Access Control}

LAA and Wi-Fi both employ similar medium access control (MAC) schemes, which categorize traffic into voice, video, background data, and priority data, in order to assign a suitable transmission opportunity (TXOP) that will guarantee QoS. In general, video and voice traffic generate smaller packets with stringent delay requirements and hence a smaller TXOP of 2 or 3 ms is assigned in LAA and 1.5 to 6 ms in Wi-Fi. In comparison, data traffic is allocated TXOPs of 8 or 10 ms in LAA and up to 6 ms in Wi-Fi. This TXOP asymmetry between LAA and Wi-Fi may lead to unfairness in accessing the channel. There are  also other specifications that create unfairness, such as the difference in initial deferment duration in LAA and Wi-Fi.

\subsection{Multi-channel Operation}
Multi-channel operation, either over wide channels as in Wi-Fi, or using carrier aggregation as in LAA, plays a key role in enhancing network capacity in unlicensed spectrum. In Wi-Fi, wide-band transmissions bond multiple narrow channels with primary and secondary channel identifiers for backward compatibility. To ensure a smooth operation, different energy and preamble detection thresholds are specified subject to whether the channel being sensed is primary or secondary.
For LAA, 3GPP specifies two modes of multi-channel operation \cite{vu2019adaptive} viz., Type A and Type B. LBT Type A offers an independent LBT back-off process to each of the aggregated 20 MHz channels, while LBT Type B runs a single back-off process only for the primary channel. Thus, LBT Type A ascertains individual channel access timing separately depending on each specific channel condition. In contrast, LBT Type B determines a common access timing for all channels solely based on the primary channel’s condition. These modes of operation may create interference on the unlicensed channel access between multi-channel LAA and multi-channel Wi-Fi, especially when there are many Wi-Fi APs on the same channel with different primary and secondary channel designations. We believe that this particular interference scenario has not been addressed in existing literature, which has focused on coexistence in a single 20 MHz channel where these issues do not arise \cite{TON,tao2015enhanced,iqbal2017impact,xu2017opportunistic,TCCN}.

\section{Data Collection Methodology}

We collected LAA and Wi-Fi measurements from January 2020 to March 2020 in various locations in Chicago, using Android phones, a custom Android App called SigCap that uses the Android API to collect cellular and Wi-Fi information from the phone and other commercial tools. The data thus collected has been processed and displayed on our website \footnote{\url{https://people.cs.uchicago.edu/~muhiqbalcr/laa/}}, and is available for download. In this section, we will describe the tools used and the data visualization methodology.

\subsection{SigCap App Development}
We developed an Android app called SigCap which is capable of simultaneously collecting Global Positioning System (GPS) data as well as cellular and Wi-Fi information using only the Android API, without requiring root access on the device. The SigCap app collects data every 10 seconds, which is the smallest interval allowed by the API in order to conserve power. 

Each data point we collect consists of the following parameters: channel number, time-stamp, location (GPS latitude and longitude), LTE cell information (Physical Cell Id (PCI)), E-UTRA Absolute Radio Frequency Channel Number (EARFCN), LTE Reference Signal Received Power (RSRP)), Wi-Fi Basic Service Set Identifiers (BSSID), Wi-Fi channel bandwidth, Wi-Fi Received Signal Strength Indicator (RSSI), and Wi-Fi operating mode such as 802.11b, 802.11n, 802.11ac. These values are extracted from the modem chip in the phone and hence conform to the relevant standard specifications.
While the SigCap app is capable of collecting data on licensed LTE channels as well as on unlicensed channels, in this paper we only consider LTE data captured on ``Band 46'' which is the designation for the unlicensed 5 GHz band.

Since all three of the major networks have deployed LAA in Chicago, we use three Android phones (two Google Pixel 3s, and a Samsung Galaxy S9) each equipped with a Subscriber Identification module (SIM) of one of the carriers, and the above app. We collected measurements while walking on sidewalks, driving a car, or riding the Chicago Transit Authority (CTA) train which run both underground and on elevated lines. We observed that the phones do not connect to the LAA secondary cell unless there is a large enough data demand: hence we initiate a large file download ($>$10 GB) before starting measurements.

\subsection{Visualization of LTE and Wi-Fi RSSI}\label{sec:heatmapMethod}

Our SigCap app measures LTE RSRP which is defined by 3GPP \cite{3gpp} as the average Reference Signal Received Power over a single Resource Element (RE) of 15 kHz wide. However, the Wi-Fi RSSI is defined by the 802.11 standard \cite{chapre2013received} as the received signal strength measured over 20 MHz\footnote{The RSSI of Wi-Fi is measured over a beacon frame which is transmitted over the 20 MHz primary channel. As per U-NII rules, the maximum transmitted power is constant over any bandwidth. Hence, we  assume that the RSSI on wider channels will be the same as in 20 MHz. There could be a slight difference due to frequency selective fading, but this will not lead to a significant difference.}. Therefore, in order to compare the received signal power of the two systems over the same bandwidth, we scale the RSRP measurement over 15 kHz to the LAA bandwidth of 20 MHz by multiplying the measured RSRP power with the number of REs in a single resource block, which is 12, and the number of physical resource blocks which is 100. This results in the LAA RSSI being defined as follows:
\begin{align}
RSSI_{(dBm)} = RSRP_{(dBm)} + 30.792.
\end{align}
The LAA and Wi-Fi RSSI are then used to create deployment and heat maps as follows. We follow the channel numbering in Fig.~\ref{fig:chdiag} to display results.

\begin{figure*}[ht]
\begin{subfigure}{.33\textwidth}
  \centering
 \includegraphics[width=5.2cm]{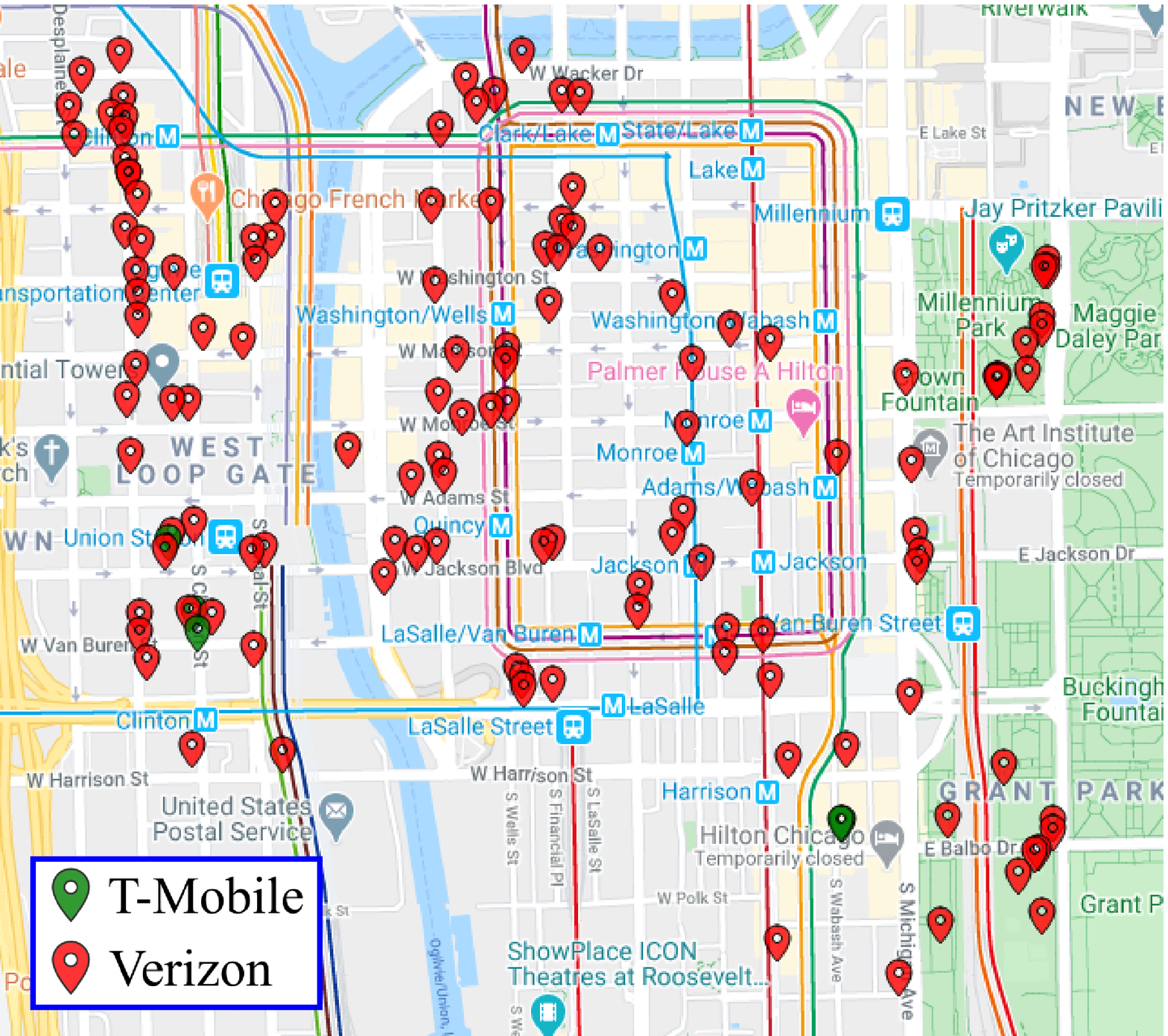}
  \caption{LAA on Channel 36}\label{fig:laa36}
\end{subfigure}
\begin{subfigure}{.33\textwidth}
  \centering
    \includegraphics[width=5.2cm]{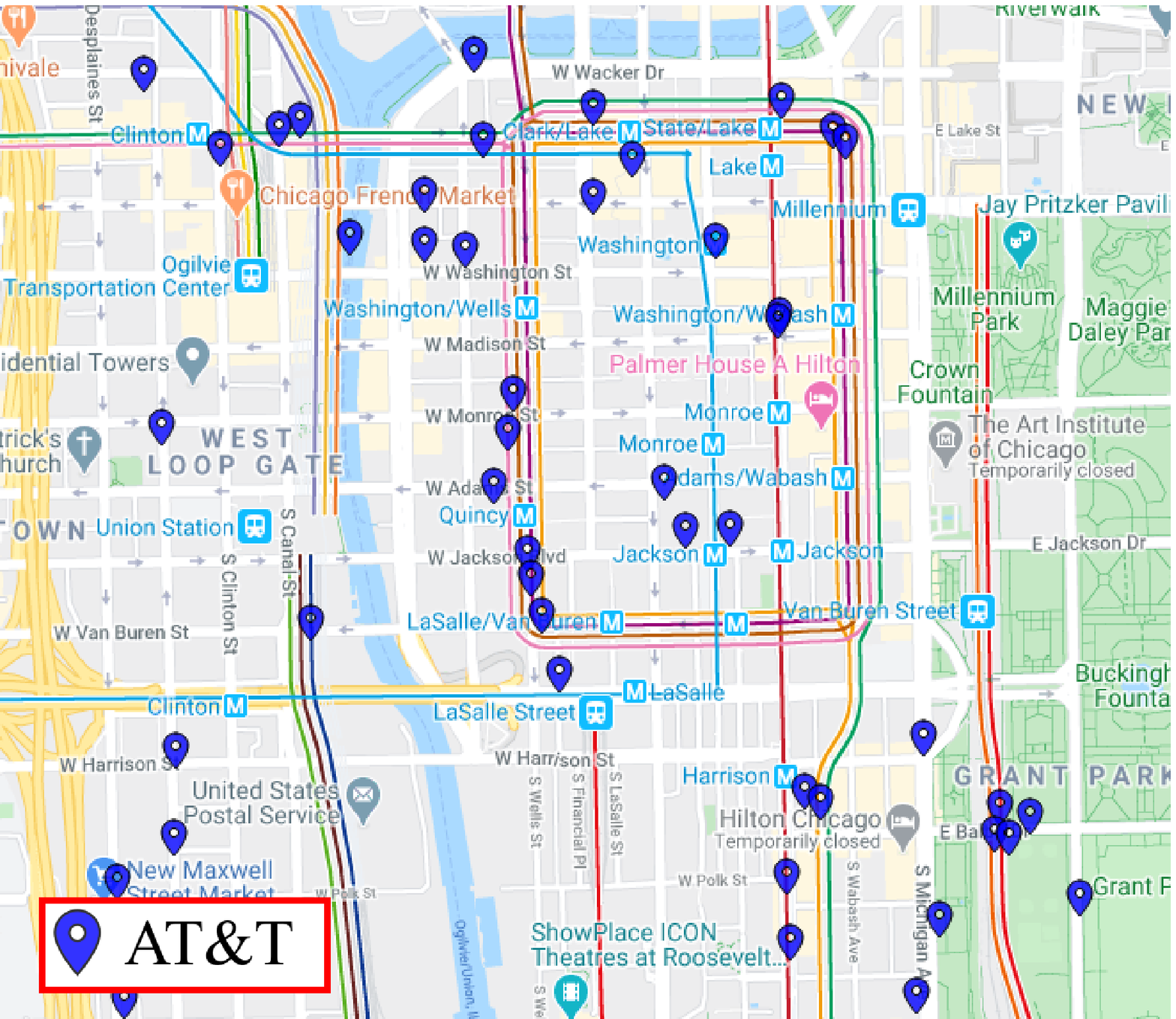}
  \caption{LAA on Channel 149}\label{fig:laa149}
\end{subfigure}
\begin{subfigure}{.33\textwidth}
  \centering
   \includegraphics[width=5.2cm]{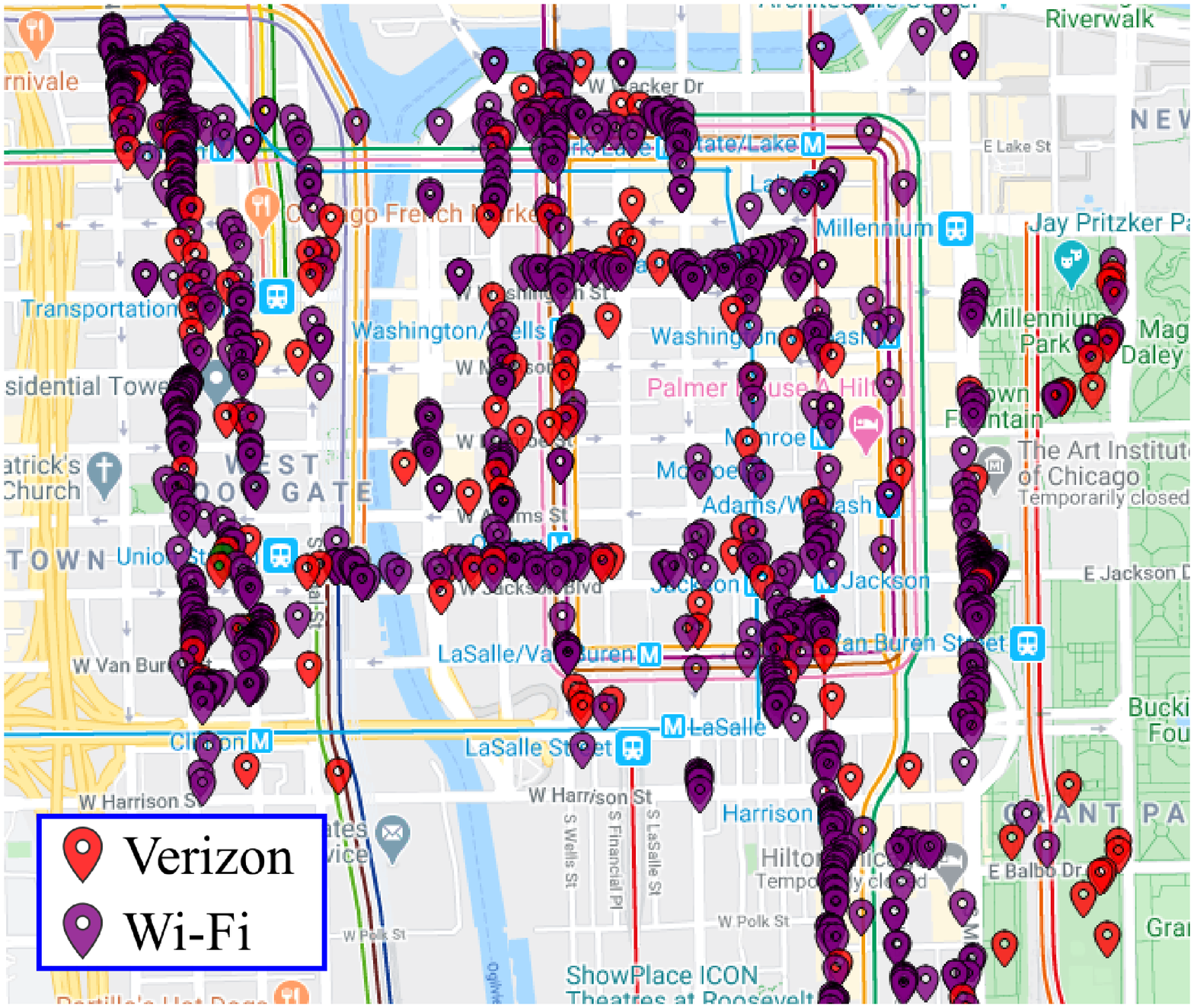}
  \caption{LAA and Wi-Fi on Channel 36}\label{fig:laawifi36}
\end{subfigure}

\caption{Deployment of LAA by AT\&T, T-Mobile and Verizon LAA and coexisting Wi-Fi on Channels 36 and 149.}
\label{fig:laawifidep}
\end{figure*}

\subsubsection{Deployment maps} The location of the LAA base stations (BSs) and Wi-Fi access points (APs) in the measurement area are unknown. In order to map the deployment, we need to be able to estimate these locations from the data we collect at the client devices. We do this as follows. For LAA, we first sort the data by the Physical Cell ID (PCI). Since PCIs are often reused, we define a reuse threshold of 1 km, \emph{i.e.}, if we observe the same PCI at a distance greater than 1 km, we treat that as a different cell \cite{pratap2018distributed}. We take the geographic centroid of the GPS coordinates of all the data corresponding to an unique PCI and use this as the approximate location of the LAA BS with that PCI.
Since accurate position or coverage of a BS is not the objective of our study, this simple method is adequate. We perform a similar operation with the Wi-Fi measurements, using BSSID as an index instead of PCI and reduce the reuse radius to 500 m. This process results in maps such as shown in Fig.~\ref{fig:laawifidep}, where each pin represents an unique LAA PCI or Wi-Fi BSSID.

\subsubsection{Heat maps} We define 11 m square grids and for each channel we average the RSSI measurements of the deployment we are interested in (LAA and/or Wi-Fi) over the grid and associate the grid-center coordinate with this average value. We process the SigCap data using text-processing scripts which output JSON files that are used by the Google Map API to create the heat map. This allows us to very quickly visualize the coverage of Wi-Fi and LAA as shown in Fig.~\ref{fig:heatmap1} where Wi-Fi deployment on varying bandwidth is also shown. LAA is only deployed on 20 MHz channels, whereas Wi-Fi can be on 20 MHz, 40 MHz or 80 MHz channels.

\subsection{Additional Tools: Network Signal Guru (NSG) and Wi-Fi analiti}
The SigCap app we developed does not require root access on the device, works on all Android devices and the collected data can be easily exported for further analysis such as creating the deployment and RSSI heat-maps described above. However, in order to obtain more detailed measurements that require root access, we used the Network Signal Guru (NSG) tool \footnote{\url{https://play.google.com/store/apps/details?id=com.qtrun.QuickTest&hl=en_US}} which can extract additional information such as throughput, latency, TXOP, signal-to-interference-and-noise ratio (SINR), block error rate (BLER), and number of allocated resource blocks (RB). We also used the Wi-Fi analiti app \footnote{\url{https://play.google.com/store/apps/details?id=com.analiti.fastest.android&hl=en_US}} to capture additional data such as the number of Wi-Fi devices on each channel and throughput. While these two commercial apps do provide additional functionality, it is laborious to export the data for analysis. Moreover, the combination of GPS location, cell ID, and RSSI that we require for efficient mapping of LTE and Wi-Fi is obtained more efficiently by our app. Due to privacy and security reasons, most of the commercial apps, including NSG and analiti, do not allow the collection of location information. Using SigCap, we can easily automate the capturing of signal and location data simultaneously. By combining the outputs of all three apps, we obtain a complete picture of LAA and Wi-Fi deployments, as presented in the rest of the paper.

\section{Overview of LAA and Wi-Fi deployments in downtown Chicago}
In this section we will present some general observations of LAA and Wi-Fi deployments as measured with our tools, followed by detailed analysis of a selected location in the next section.

\subsection{Overview of LAA deployments in Chicago}
We collected data from January to March 2020 using all the tools described in Section III in different areas of Chicago such as the Loop, South Loop, and River North areas of downtown, and three major University campuses: University of Chicago (UChicago), University of Illinois at Chicago (UIC) and the Illinois Institute of Technology (IIT)). All three major cellular operators, AT\&T, T-Mobile and Verizon, have extensive LAA deployments in these locations. In total, we collected 5114 data points representing 557 unique LAA PCIs and 10639 unique Wi-Fi BSSIDs in the vicinity of the observed LAA deployments (we only report Wi-Fi measurements in locations where we also observe at least one LAA PCI). We observe that AT\&T and Verizon's LAA cells are primarily deployed around the densely populated Loop area (as shown on Fig.~\ref{fig:laa36} and \ref{fig:laa149}), while the T-Mobile deployments are in the less dense residential areas and on the IIT campus. We also note that recent measurements not reported here indicate that these deployments have been rapidly expanding in recent months.

All LAA deployments we measured were based on Release 13, which means that all downlink transmissions are over the unlicensed band while all uplink transmissions are over the licensed band.  Generally, if the SINR is high, we observe maximum of three aggregated 20 MHz unlicensed channels\footnote{Number of aggregated channels also depends on the mobile device capability. For example: iPhone 10S only supports aggregating two unlicensed channels, compared to the phones we use in our experiments, Samsung Galaxy S9 and Google Pixel 3, which support aggregation on three unlicensed channels}. During high mobility or when the SINR is low, we occasionally observe aggregation over 1 or 2 unlicensed channels.

\subsection{Overview of Wi-Fi APs in the vicinity of LAA}\label{sec:wifioverview}

As mentioned previously, in this paper we only report Wi-Fi measurements in the vicinity of at least one LAA PCI. Table~\ref{tab:laawifidepstat} summarizes our observations on the channel assignment statistics of Wi-Fi and LAA in Chicago. The table shows that the number of overlapping Wi-Fi APs in each channel does not add up to the total number of unique BSSID (10639). This is because in each row we count all APs that overlap with LAA's 20 MHz channel including wider channels, e.g., a Wi-Fi AP operating on Channels 38 (40 MHz) or 42 (80 MHz) overlaps with Channel 36, and is therefore counted as overlapping with Channel 36. 

Since we have not observed any LAA deployments on the U-NII-2 channels, we do not include any Wi-Fi measurements in U-NII-2. The absence of U-NII-2 LAA deployments, we believe, is due to the mandatory DFS requirement in that band. According to our measurements, Verizon and T-Mobile LAA mostly use U-NII-1 Channels 36, 40, and 44. Fig.~\ref{fig:laa36}\footnote{The deployment map of Channel 36, 40 and 44 are similar so we have shown only on Channel 36.}, shows the LAA deployments by T-Mobile and Verizon on Channel 36, 40 and 44 in the Loop area. AT\&T deploys LAA primarily in U-NII-3 on Channel 149, 153, 157, 161 and 165 as shown in Fig.~\ref{fig:laa149}\footnote{The deployment map of Channel 149, 153 and 157 are similar so we have shown only on Channel 149.}.

As shown in Table~\ref{tab:laawifidepstat} and Fig.~\ref{fig:laawifi36}, Wi-Fi is deployed densely in the vicinity of LAA base-stations, potentially leading to coexistence problems as increasing numbers of new mobile devices capable of Band 46 operation begin to penetrate the market.
Using the NSG app, we observe that the SINR of LAA increases in the vicinity of certain street lamps, which leads us to believe that most LAA is deployed outdoors, while the Wi-Fi APs are deployed mostly by shops and hence are indoors. However, client devices capable of LAA and Wi-Fi could be both indoors and outdoors, which combined with the dense deployment, may lead to interesting coexistence scenarios that have not been comprehensively studied in existing literature. Additionally, we identified a coexistence scenario where LAA aggregated over three 20 MHz channels will coexist with 80 MHz Wi-Fi: a research problem that has not been studied in depth in the research literature.
\begin{figure*}[htb]
\begin{subfigure}{.33\textwidth}
  \centering
 \includegraphics[width=5.2cm, height=5cm]{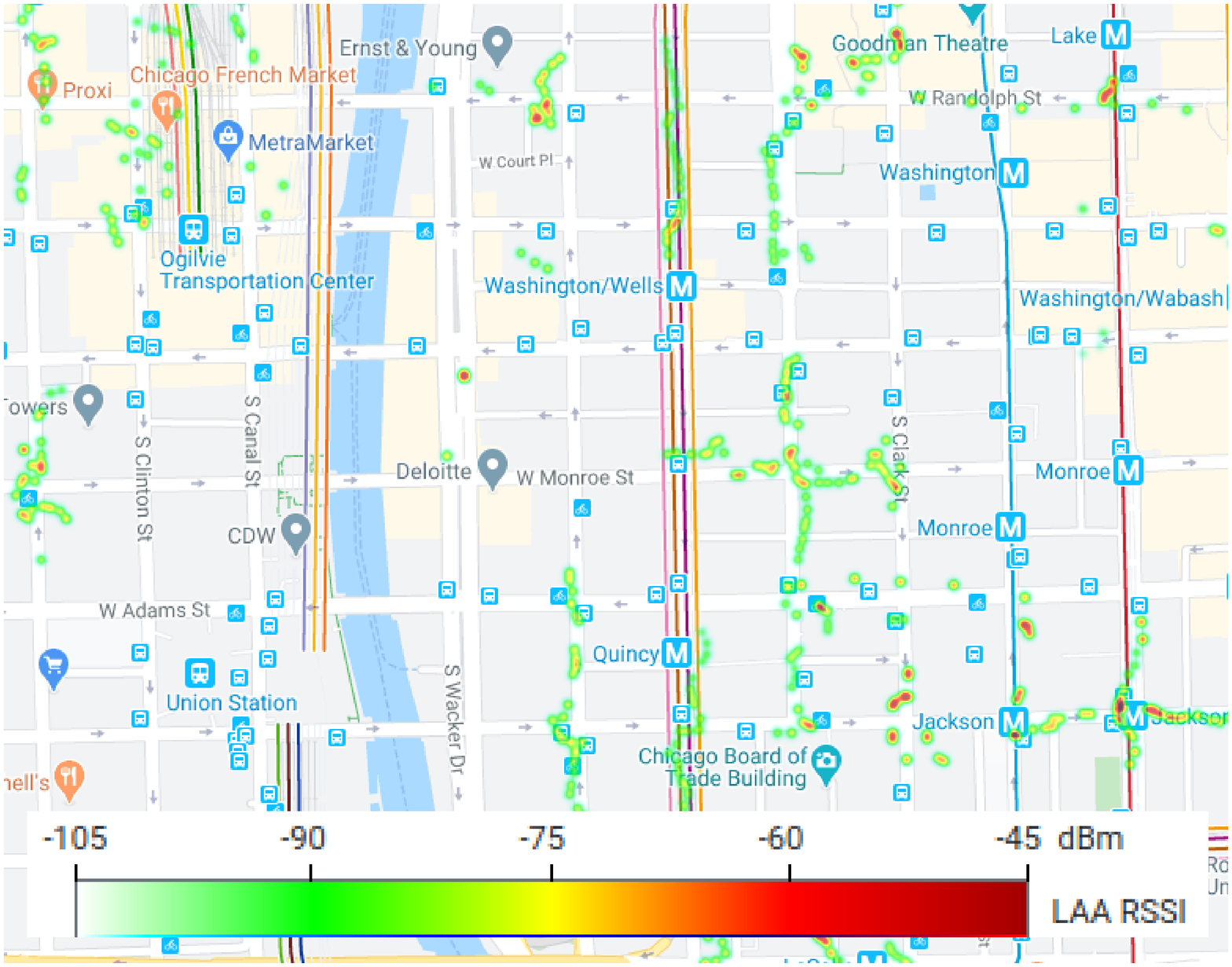}
  \caption{AT\&T LAA RSSI on Channel 157}\label{watt1}
\end{subfigure}
\begin{subfigure}{.33\textwidth}
  \centering
    \includegraphics[width=5.2cm, height=5cm]{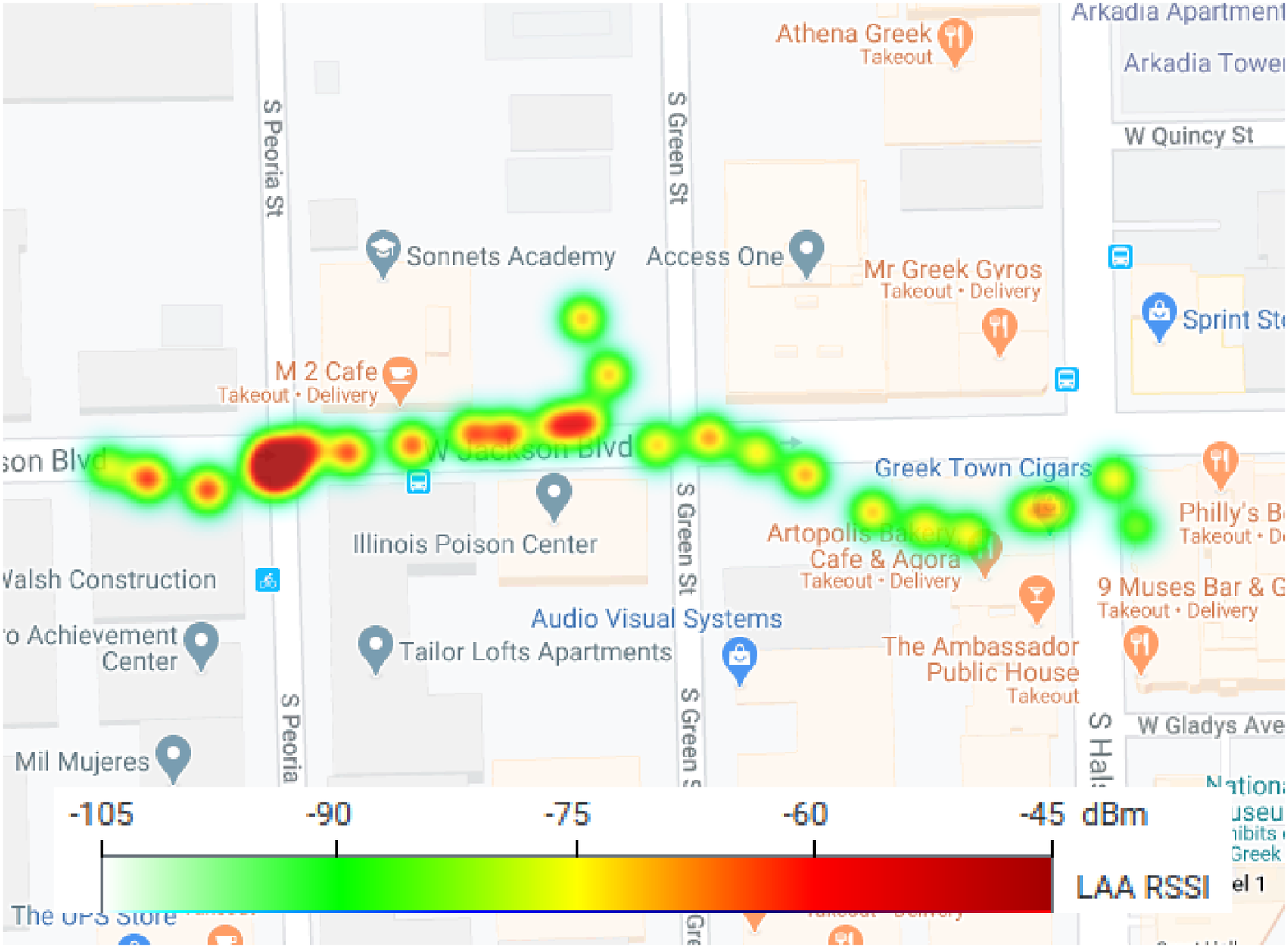}
  \caption{T-Mobile LAA RSSI on Channel 36}\label{wtmobile}
\end{subfigure}
\begin{subfigure}{.33\textwidth}
  \centering
   \includegraphics[width=5.2cm, height=5cm]{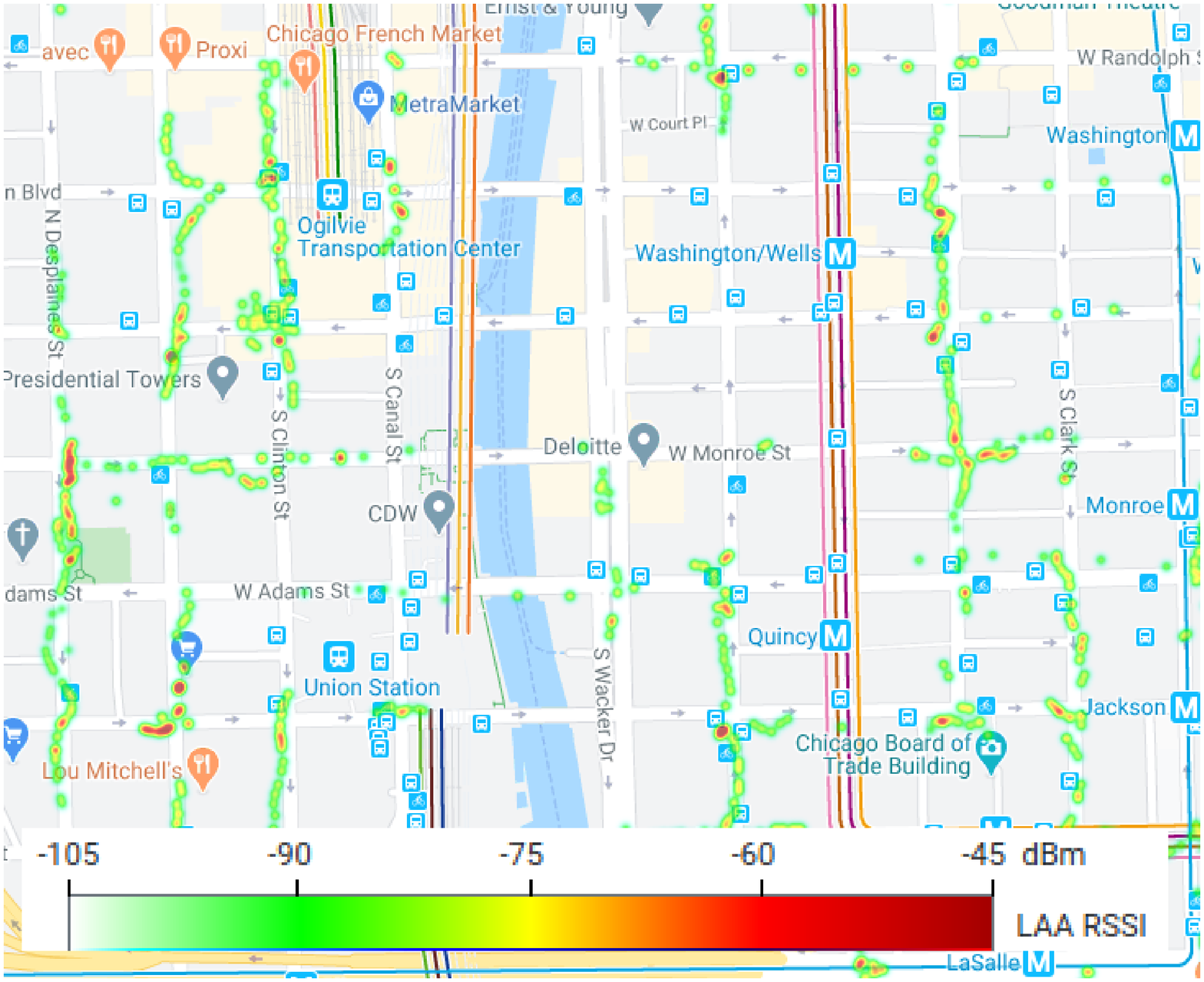}
  \caption{Verizon LAA RSSI on Channel 36}\label{wverizon}
\end{subfigure}
\bigskip

\begin{subfigure}{.33\textwidth}
  \centering
 \includegraphics[width=5.2cm, height=5cm]{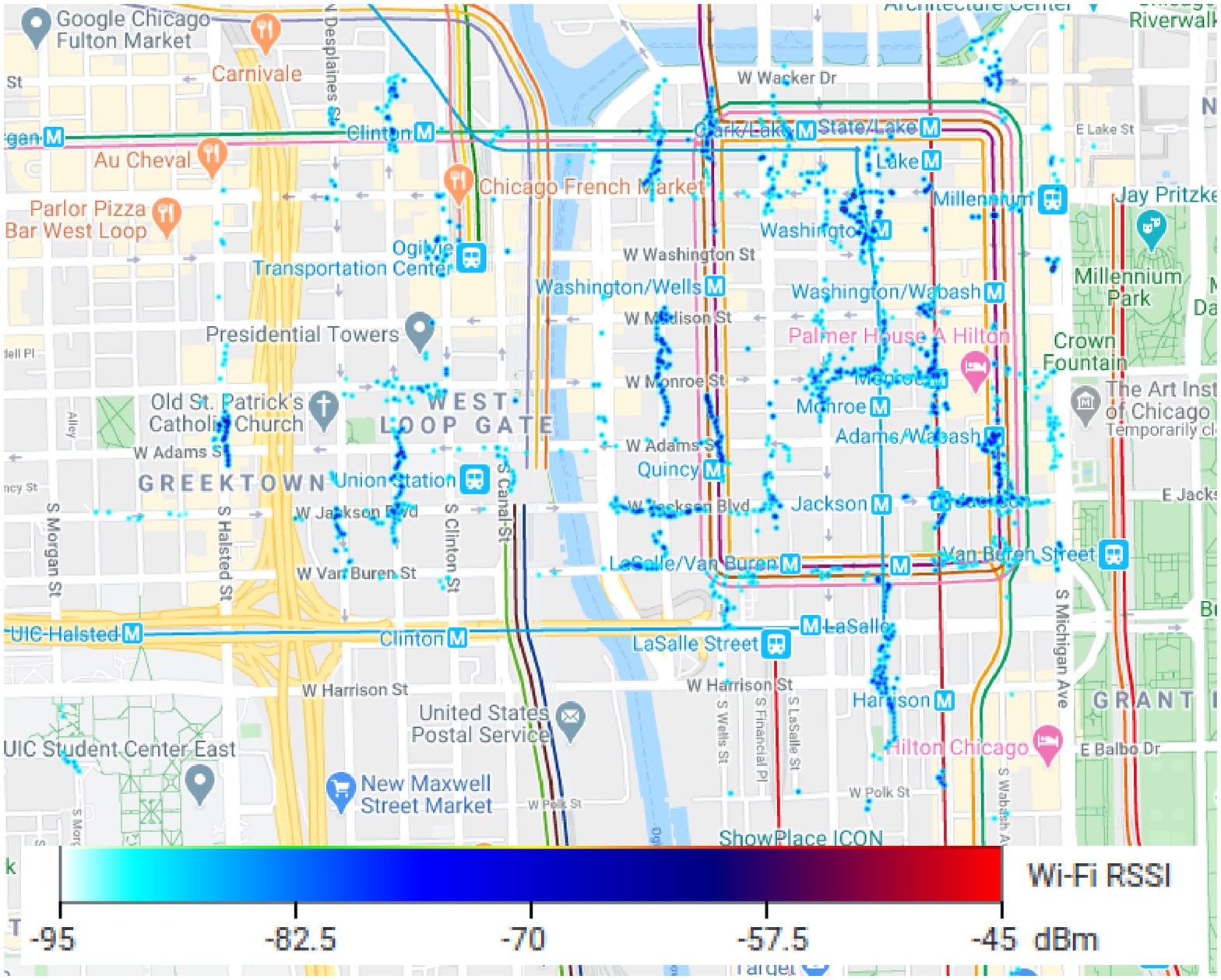}
  \caption{Wi-Fi RSSI heat-map, 20 MHz channels}\label{w20}
\end{subfigure}
\begin{subfigure}{.33\textwidth}
  \centering
    \includegraphics[width=5.2cm, height=5cm]{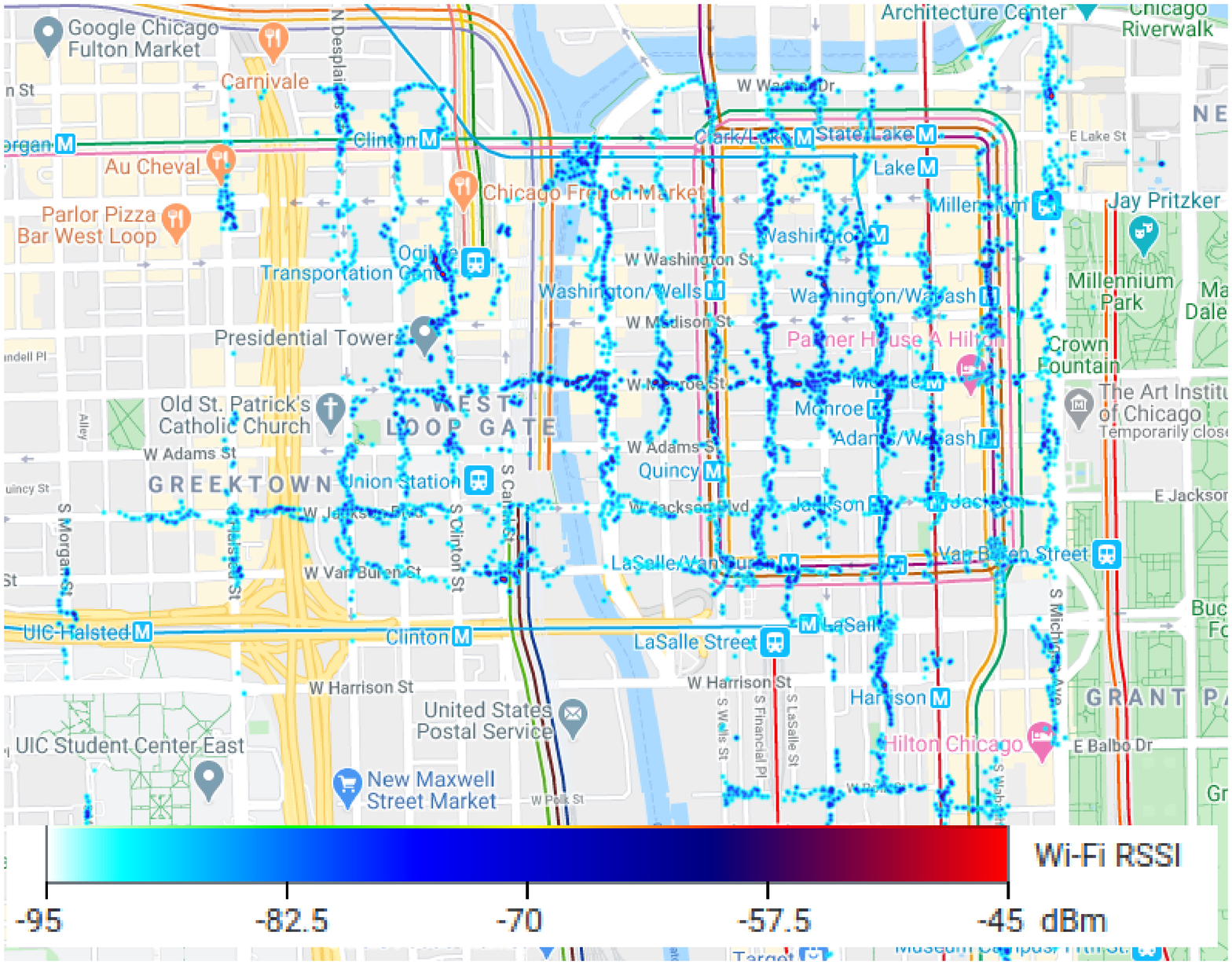}
  \caption{Wi-Fi RSSI heat-map, 40 MHz channels}\label{w40}
\end{subfigure}
\begin{subfigure}{.33\textwidth}
  \centering
   \includegraphics[width=5.2cm, height=5cm]{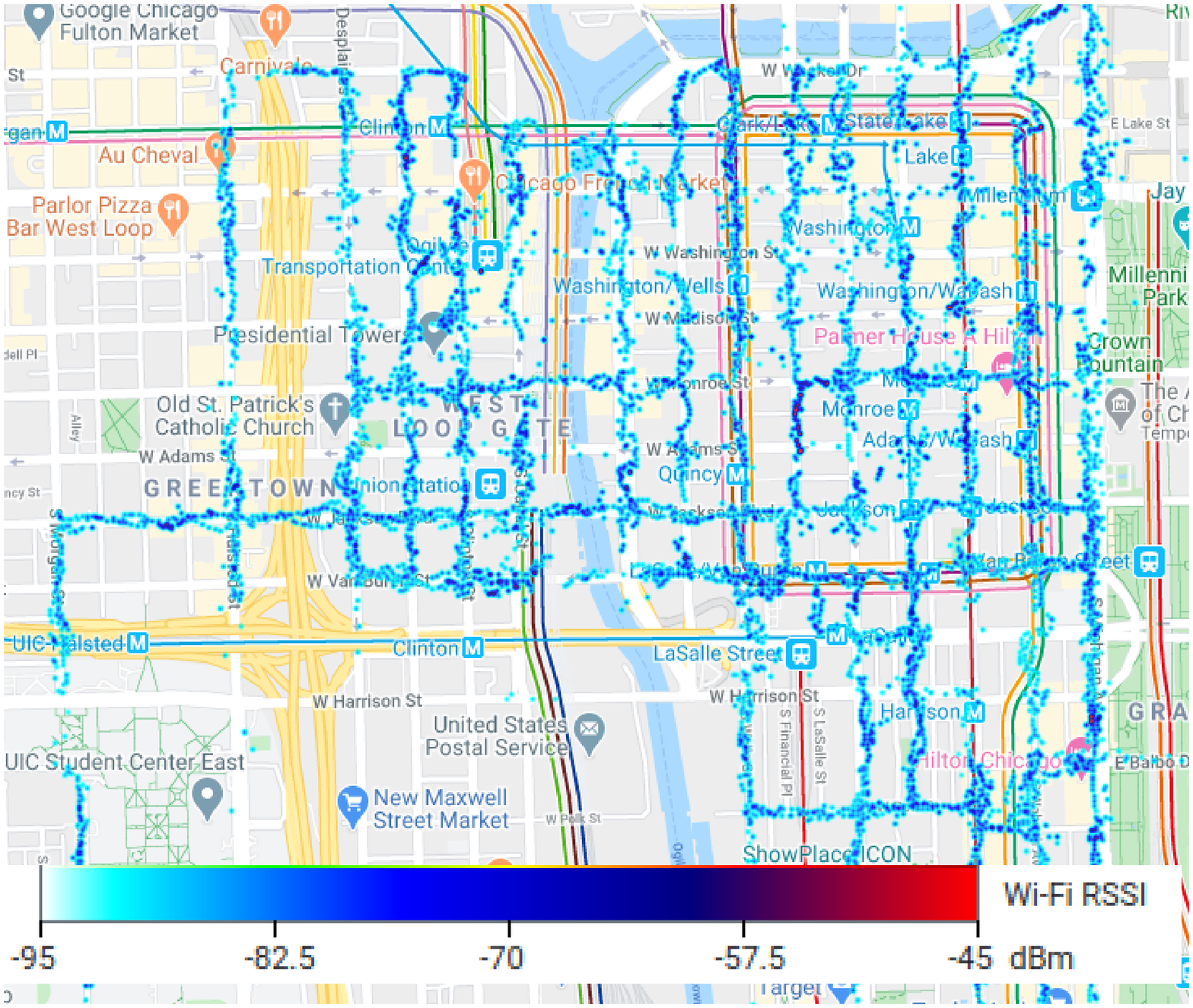}
  \caption{Wi-Fi RSSI heat-map, 80 MHz channels}\label{w80}
\end{subfigure}
\bigskip

\begin{subfigure}{.33\textwidth}
  \centering
 \includegraphics[width=5.2cm, height=5cm]{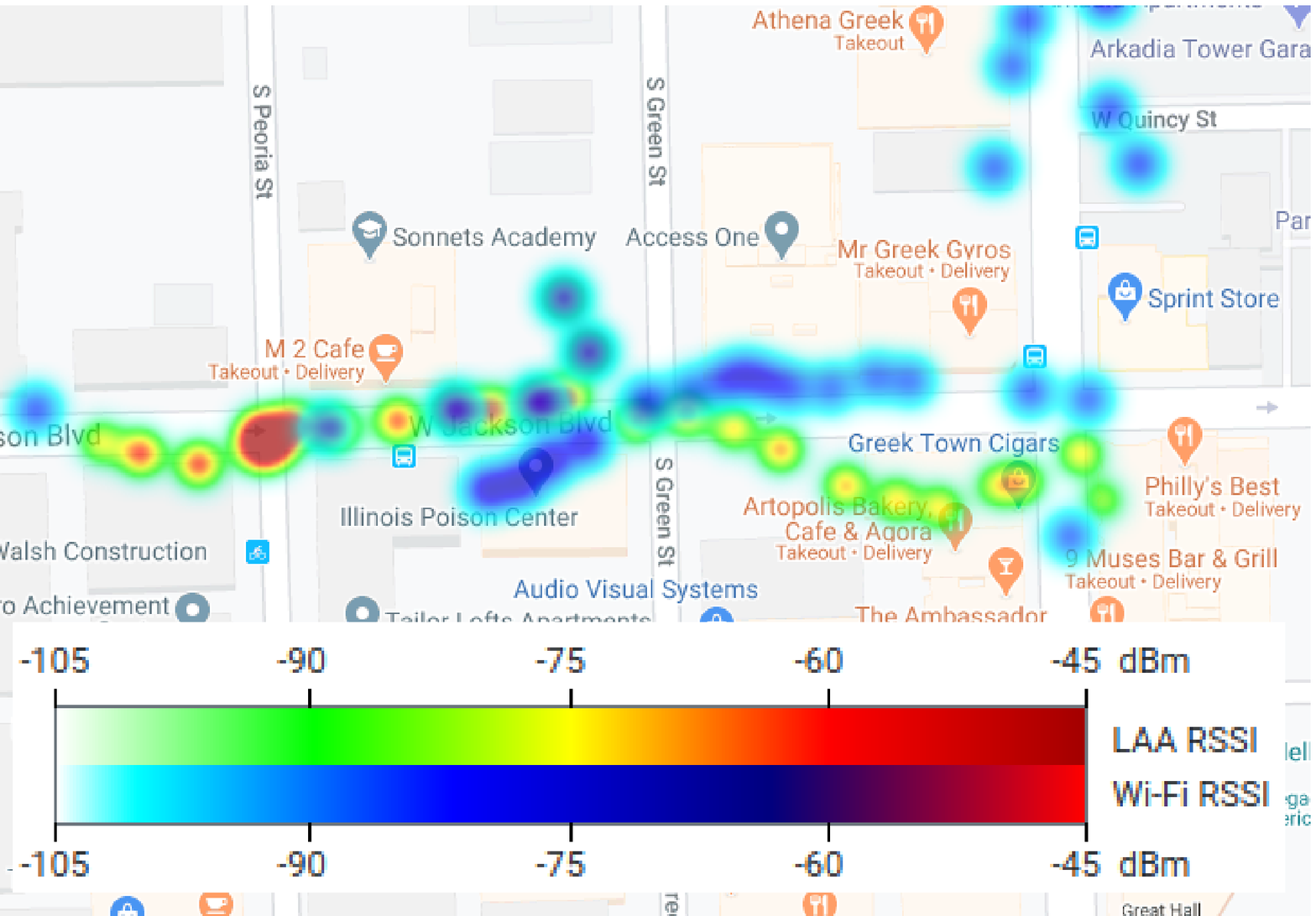}
  \caption{LAA and Wi-Fi RSSI, 20 MHz channels}\label{tw20}
\end{subfigure}
\begin{subfigure}{.33\textwidth}
  \centering
   \includegraphics[width=5.2cm, height=5cm]{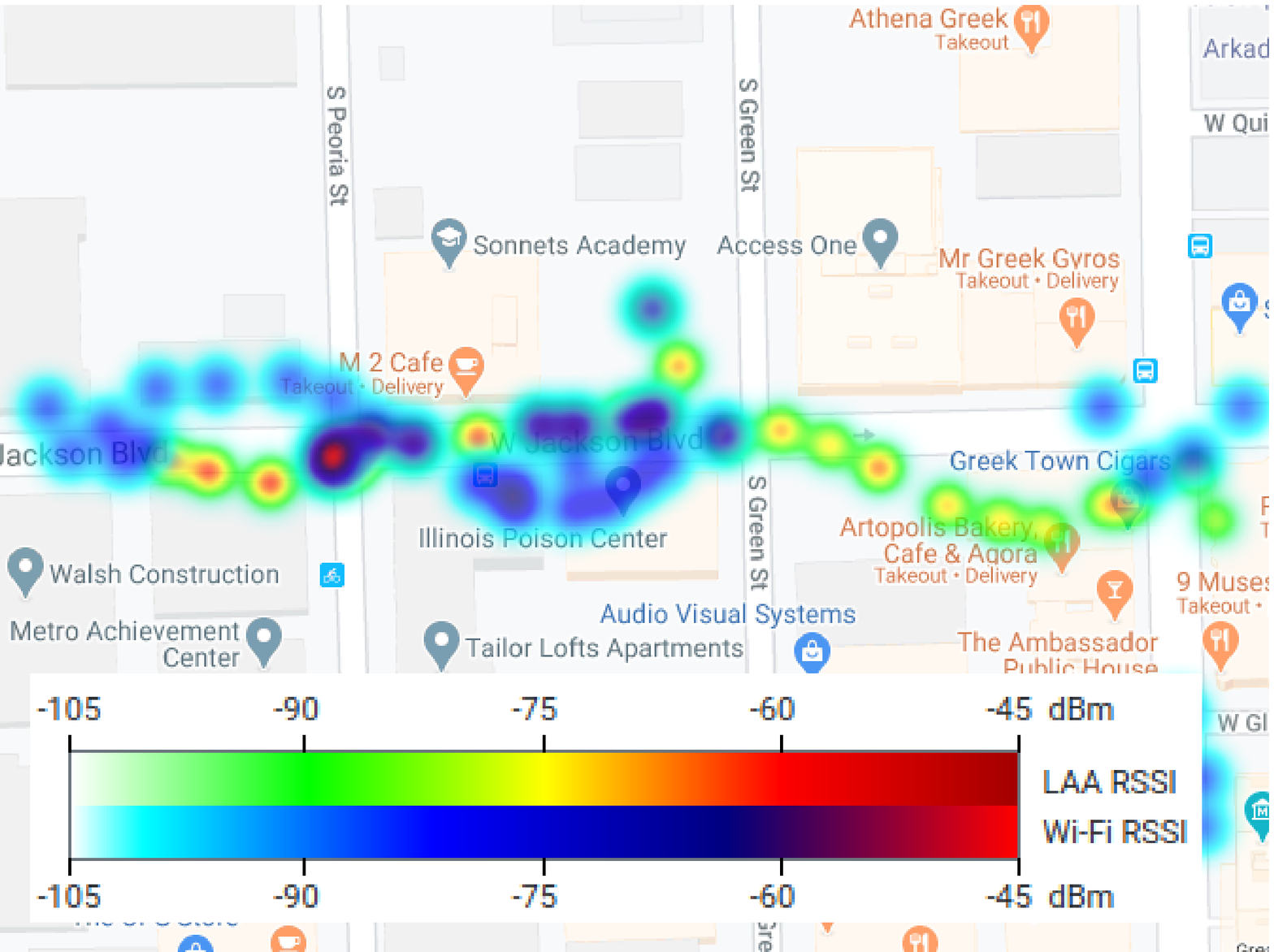}
  \caption{LAA and Wi-Fi RSSI, 40 MHz channels}\label{tw40}
\end{subfigure}
\begin{subfigure}{.33\textwidth}
  \centering
   \includegraphics[width=5.2cm, height=5cm]{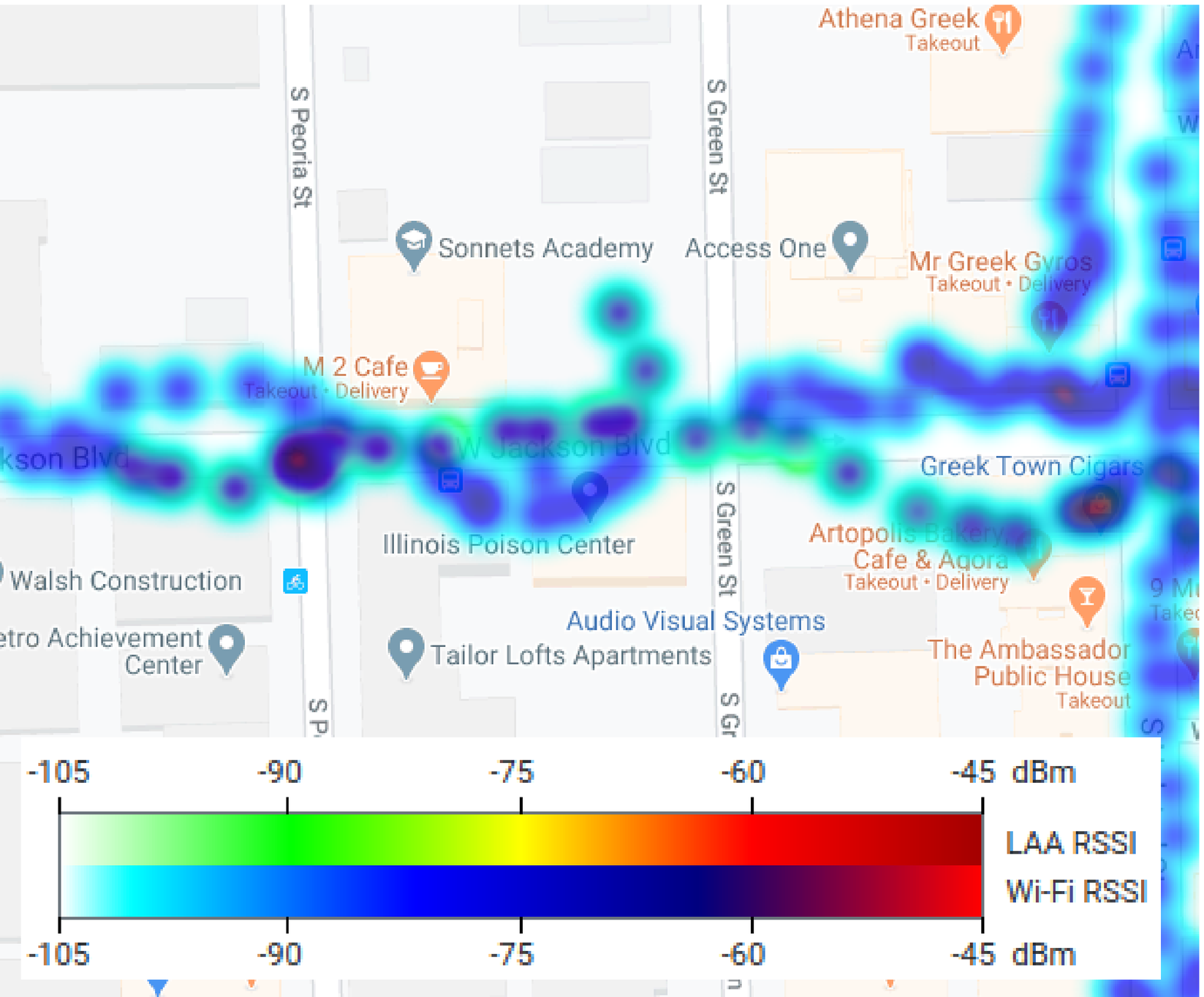}
  \caption{LAA and Wi-Fi RSSI, 80 MHz channels}\label{tw80}
\end{subfigure}
\caption{LAA and Wi-Fi RSSI heat-maps on 20 MHz, 40 MHz and 80 MHz channels.}
\label{fig:heatmap1}
\end{figure*}

\begin{table}
\begin{subtable}{0.45\textwidth}
\small
\begin{tabular}{| *{20}{c|} }
\hline
\cellcolor{Gray} & \multicolumn{3}{c|}{\cellcolor{Gray} \textbf{LAA}} & \cellcolor{Gray} \textbf{Over-} \\
\cellcolor{Gray} \textbf{Channels} & \multicolumn{3}{c|}{\cellcolor{Gray} \textbf{Deployment}} & \cellcolor{Gray} \textbf{lapping} \\
\cline{2-4} \cellcolor{Gray} &{\cellcolor{Gray} \textbf{AT\&T}}&{\cellcolor{Gray} \textbf{T-Mobile}} &{\cellcolor{Gray} \textbf{Verizon}} & \cellcolor{Gray} \textbf{Wi-Fi}\\
\hline
\hline
Channel 36  &   4  &   49  &   209 &   3089     \\
\hline
Channel 40   &  5 &   38    &   222  &   2750   \\
\hline
Channel 44   &  4 &   38    &   205  &   3009   \\
\hline
Channel 48   &  0 &   0    &   0  &   3189   \\
\hline
Channel 149   &  104 &   0    &   2  &   2673   \\
\hline
Channel 153   &  119 &   1    &   4  &   2569   \\
\hline
Channel 157   &  131 &   1    &   22  &   2648   \\
\hline
Channel 161  &  38 &   0    &   25  &   2601   \\
\hline
Channel 165   &  27 &   0    &   20  &   105   \\
\hline
\end{tabular}
\caption{LAA and Wi-Fi Deployment Statistics.}
\label{tab:laawifidepstat}
\end{subtable}

\bigskip

\begin{subtable}{0.43\textwidth}
\footnotesize
\begin{tabular}{| *{20}{c|} }
\hline
\cellcolor{Gray} & \cellcolor{Gray} & \multicolumn{3}{c|}{\cellcolor{Gray} \textbf{Wi-Fi}} \\
\cline{3-5}
\cellcolor{Gray} \textbf{RSSI Metrics} & \cellcolor{Gray} \textbf{LAA} & \cellcolor{Gray} \textbf{20 MHz} & \cellcolor{Gray} \textbf{40 MHz} & \cellcolor{Gray} \textbf{80 MHz} \\
\hline
\hline
Mean (dBm) & -84.49 &   -82.19  &   -82.87 &   -83.65     \\
\hline
Variance & 63.92 & 46.41 & 34.55 & 19.3 \\
\hline
Range (dBm) & [-105, -45] & [-95, -47] & [-95, -56] & [-95, -51]   \\
\hline
\end{tabular}
\caption{LAA and Wi-Fi RSSI Statistics.}
\label{tab:laawifirssistat}
\end{subtable}

\caption{LAA and Wi-Fi Deployment and RSSI Statistics}
\end{table}

\subsection{RSSI Heat-map of LAA and Wi-Fi}

We measured Wi-Fi and LAA RSSI while walking outdoors using the methodology described in Sec.~\ref{sec:heatmapMethod}. Table \ref{tab:laawifirssistat} summarizes the statistics of LAA and Wi-Fi RSSI measurements. The maximum RSSI observed on LAA is higher than the maximum on Wi-Fi corroborating our conjecture that most LAA BSs are deployed outdoors while Wi-Fi is deployed indoors. The Wi-Fi RSSI data exhibits a lower variance: this is because the Wi-Fi RSSI is averaged over all BSSIDs and in general the Wi-Fi deployment is denser than LAA: hence the signal level outdoors due to Wi-Fi exhibits more uniformity than LAA.

Fig.~\ref{watt1}, \ref{wtmobile} and \ref{wverizon} show the RSSI heat-maps for AT\&T, T-Mobile and Verizon. Since T-Mobile's LAA deployment in the downtown area is sparser than AT\&T's and Verizon's, Fig.~\ref{wtmobile} shows a zoomed-in view of the region where T-Mobile is deployed. Fig.~\ref{w20}, \ref{w40}, and \ref{w80} show the heat-maps of Wi-Fi APs on 20 MHz, 40 MHz and 80 MHz channels. We see that the majority of deployed Wi-Fi APs today implement 80 MHz channels and there are only a handful of legacy 20 MHz APs. Fig.~\ref{tw20}, \ref{tw40} and \ref{tw80} show LAA and Wi-Fi RSSI on the same map, demonstrating that the two deployments have significant overlap and hence the potential of mutual interference is high.

\begin{figure*}
\centering
\begin{minipage}{0.45\textwidth}
\centering
\includegraphics[width=\textwidth]{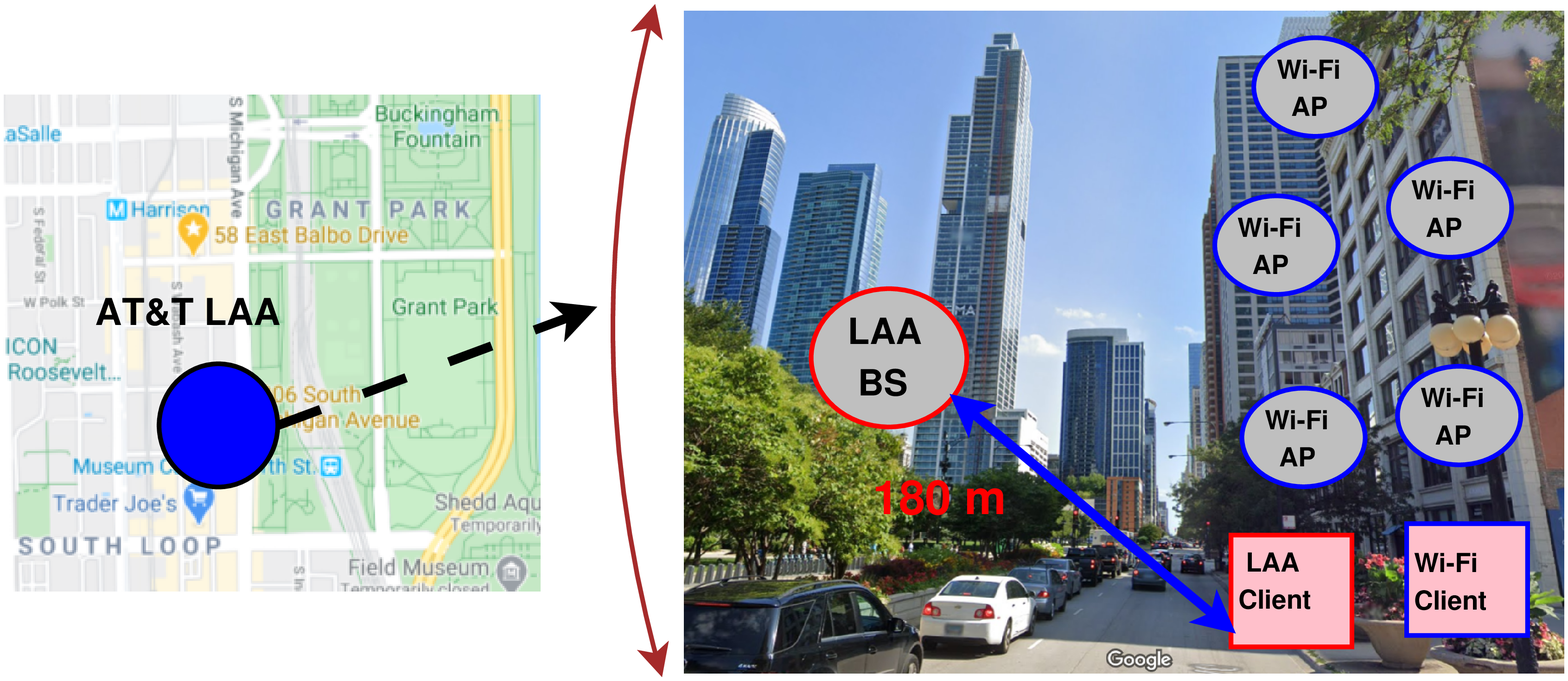}
\subcaption*{(a) Experiment location in downtown Chicago}
\label{downtown}
\end{minipage}
\qquad
\begin{minipage}{0.45\textwidth}
\centering
\captionsetup{type=table} 
\begin{tabular}{| *{20}{c|} }
    \hline
\cellcolor{Gray} & \multicolumn{3}{c|}{\cellcolor{Gray} \textbf{W/L Coexistence}} & \multicolumn{3}{c|}{\cellcolor{Gray} \textbf{W/W Coexistence}}  \\
\cline{2-7} 
\cellcolor{Gray} \textbf{Traffics} &  \cellcolor{Gray} \textbf{149} & \cellcolor{Gray} \textbf{153} & \cellcolor{Gray} \textbf{157}  &  \cellcolor{Gray} \textbf{149} & \cellcolor{Gray} \textbf{153} & \cellcolor{Gray} \textbf{157} \\
    \hline
    \hline
D  &   4+ &  3+ & 9+ &  12+ & 11+ & 31+  \\
    \hline
D+V  &   3+ &  2+ & 6+ & 10+ & 7+ &  15+ \\
    \hline
S   &   13+  &  10+ & 31+ &  19+ & 20+ &  45+ \\
    \hline
D+S &   2+ &  1+ & 5+ &  9+ & 7+ & 16+ \\
    \hline
V &   5+ &  3+ & 10+ & 21+ & 14+ &  36+  \\
\hline
\end{tabular}
\subcaption*{(b) Number of Wi-Fi associated devices (on the LAA-occupied channels) in Wi-Fi/Wi-Fi (W/W) and Wi-Fi/LAA (W/L) Coexistence}
\end{minipage}
\caption{Experiment Location for LAA and Wi-Fi coexistence.}
\label{figuretabledep}

\end{figure*}

\section{Representative measurements of coexisting LAA and Wi-Fi}\label{is}

While we made measurements with all the carriers in multiple locations, due to space limitations, we present detailed measurements and discussions of coexistence of LAA and Wi-Fi in one representative location where AT\&T has deployed LAA. We use the NSG tool to calculate the TXOP, SINR, RBs and throughput of LAA with all results averaged over 10 measurements in the same location. Similarly, the analiti tool is used to characterize the Wi-Fi deployment in terms of number of Wi-Fi clients per channel, and average delay.  

\begin{figure*}[htb]
\begin{subfigure}{.5\textwidth}
  \centering
 \includegraphics[width=3.4in]{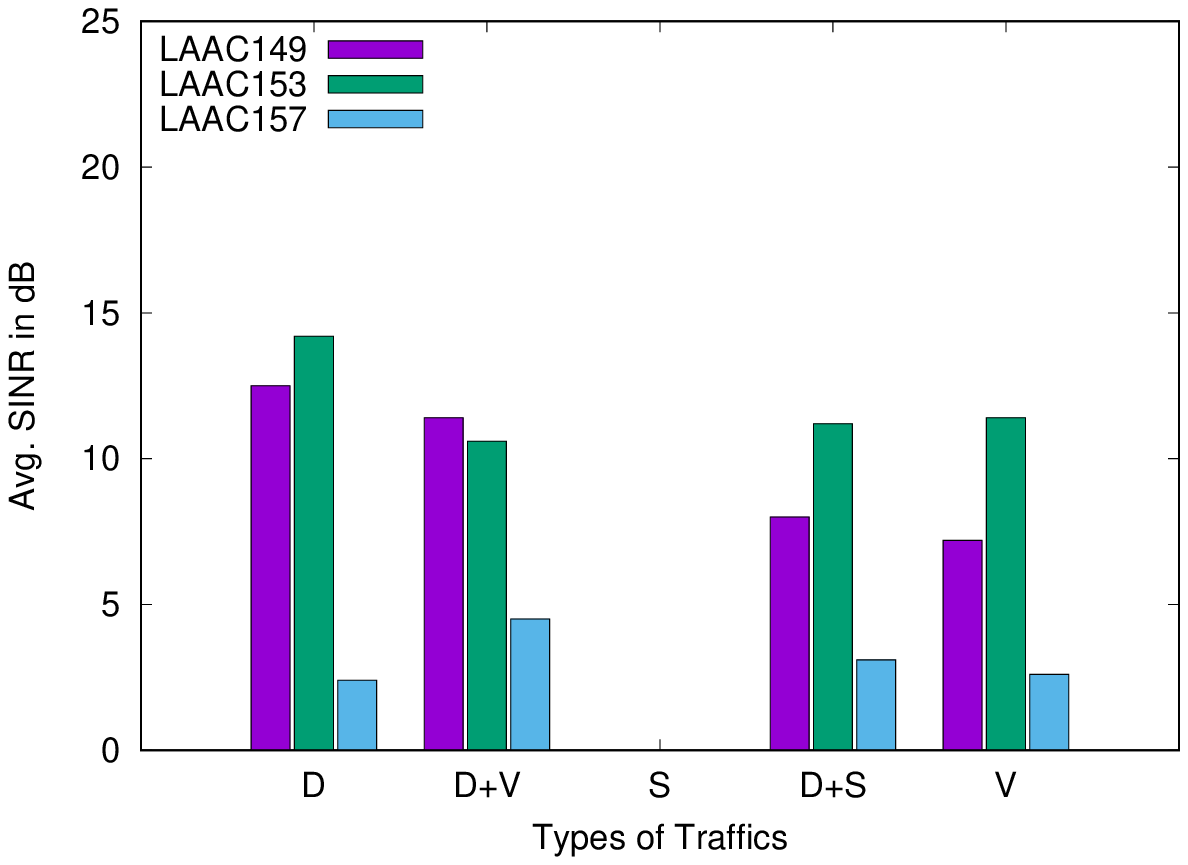}
  \caption{AT\&T Avg. SINR}\label{asinr}
\end{subfigure}
\begin{subfigure}{.5\textwidth}
  \centering
    \includegraphics[width=3.4in]{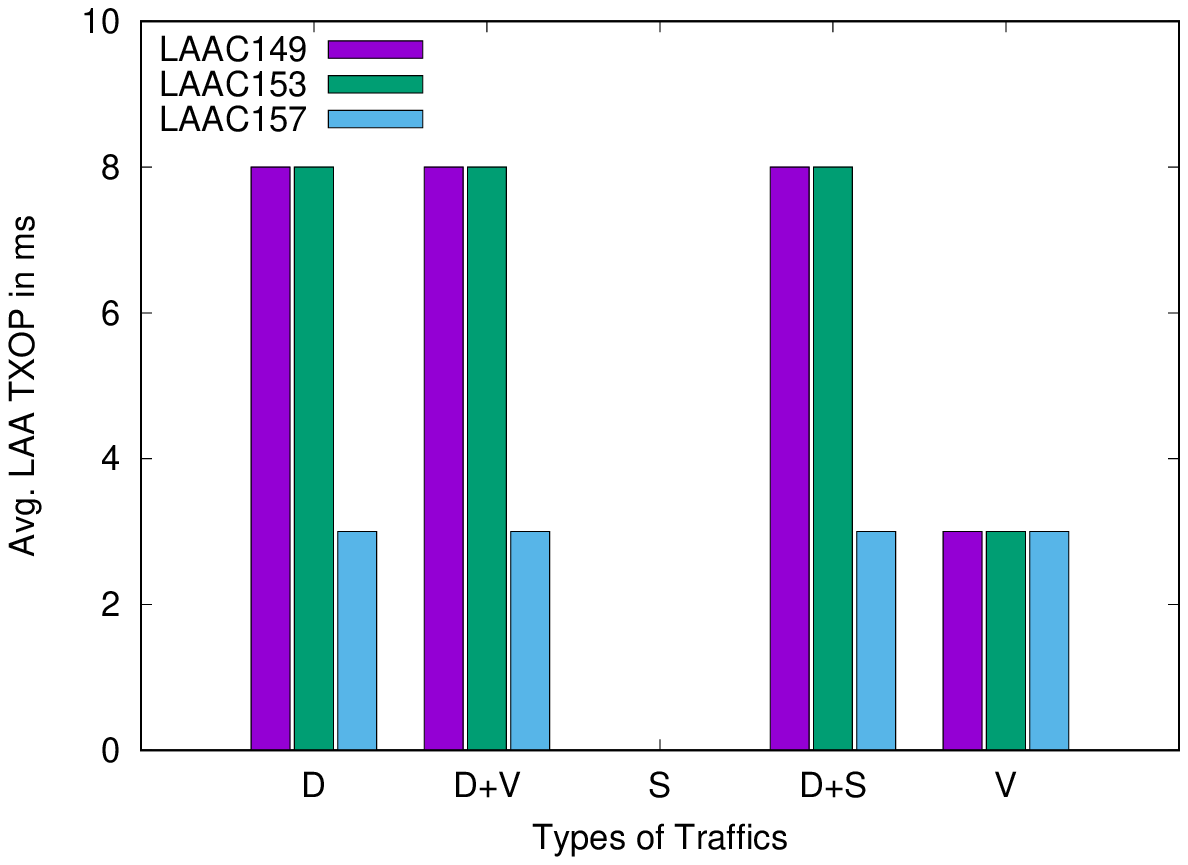}
  \caption{AT\&T Avg. TXOP}\label{atxop}
\end{subfigure}

\bigskip

\begin{subfigure}{.33\textwidth}
  \centering
   \includegraphics[width=2.4in]{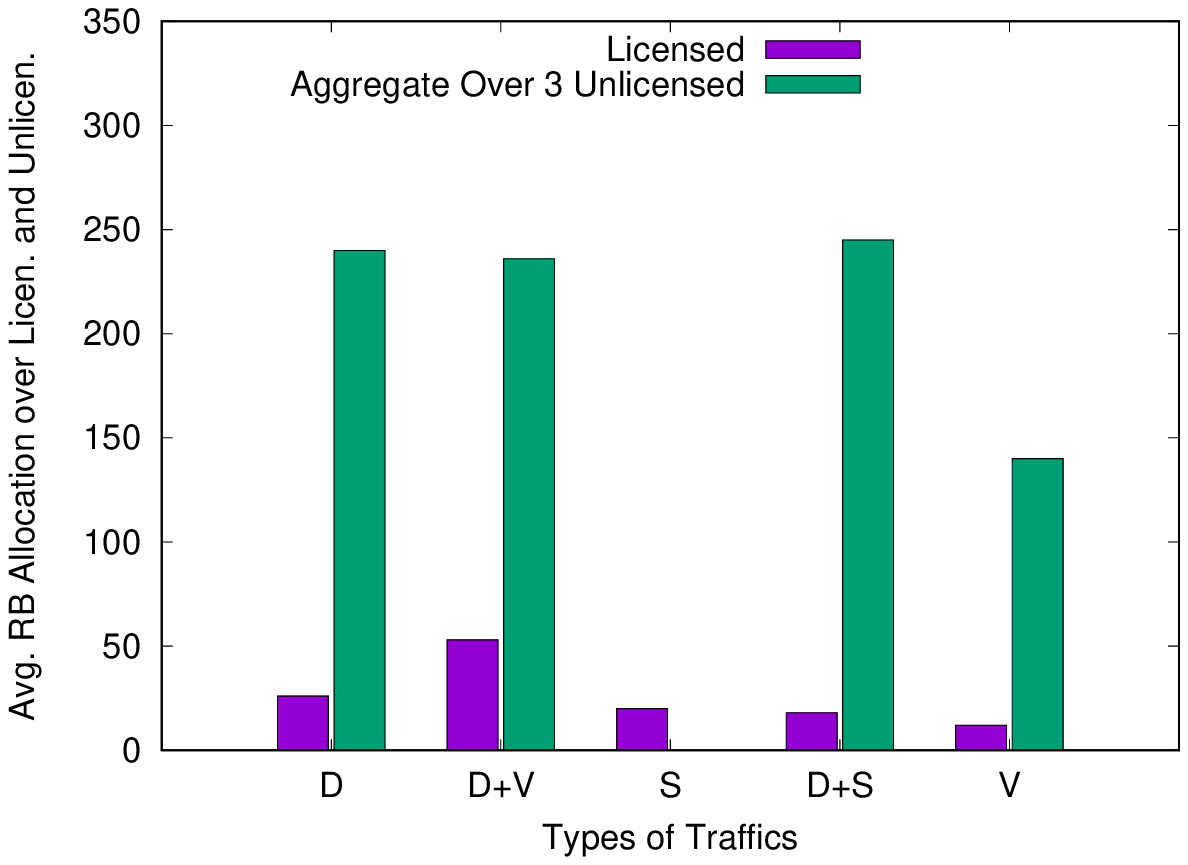}
  \caption{AT\&T Avg. RB}\label{arb}
\end{subfigure}
\begin{subfigure}{.33\textwidth}
  \centering
   \includegraphics[width=2.3in]{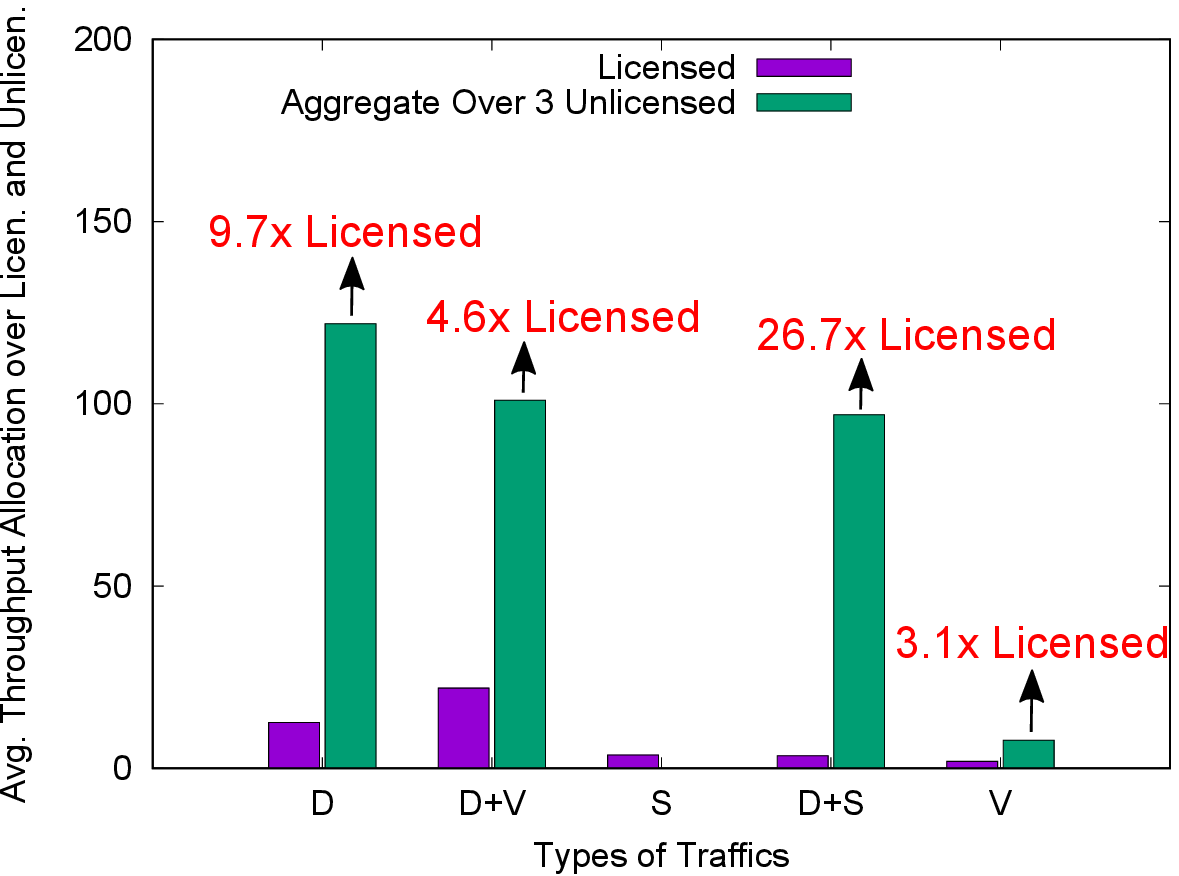}
  \caption{AT\&T Avg. Throughput}\label{ath}
\end{subfigure}
\begin{subfigure}{.33\textwidth}
  \centering
   \includegraphics[width=2.4in]{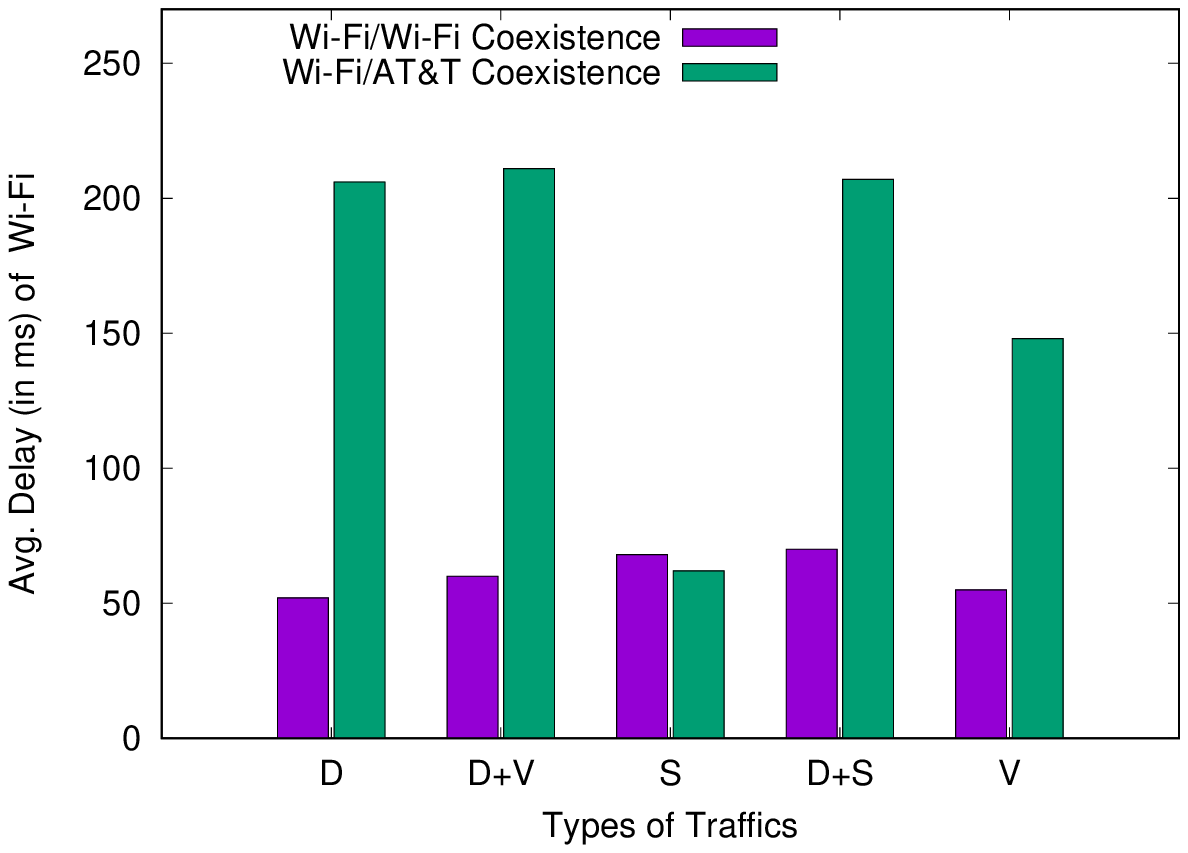}
  \caption{Avg. Delay of Wi-Fi under W/W Vs W/L}\label{delay}
\end{subfigure}

\caption{Average SINR, TXOP, RB, Throughput and Delay for different traffic types at AT\&T LAA BS.}
\label{fig:laatxop}
\end{figure*}

\subsection{Location and traffic description}
Fig.~\ref{figuretabledep}a shows the measurement location and environment. AT\&T LAA is deployed on the street lamp at 1006 S Michigan Ave, Chicago, IL 60605. This particular BS operates on Channel 149, 153 and 157 for a maximum 60 MHz of unlicensed spectrum in addition to the 15 MHz licensed primary channel, for a total maximum of 75 MHz bandwidth and 375 RBs. During the experiment, the number of Wi-Fi APs coexisting on LAA channel 149, 153 and 157 is at least 6, 5 and 8 respectively. Since most of the Wi-Fi APs we detect are deployed by shops and other establishments and are password protected, we cannot connect to these networks. However, using our personal Xfinity credential, we can connect to any Xfinity Wi-Fi APs for this experiment's purpose. We chose a location in Chicago downtown where Xfinity Wi-Fi APs were deployed close to the LAA location: using our phones to connect to the APs enabled us to access Wi-Fi information such as operating channels, the number of clients connected, operating bandwidth, operating Wi-Fi mode.

In order to study both LAA and Wi-Fi resource allocation and throughput for different traffic types, we place a LAA and Wi-Fi client next to each other and initiate downloads for the following traffic types:
\begin{itemize}
    \item \textbf{Data (D):} Pure data traffic is generated by downloading a large YUV dataset ($>$10 GB) from Derf Test Media Collection \footnote{\url{https://media.xiph.org/video/derf}}.
    \item \textbf{Video (V):} A Youtube video is downloaded, with a resolution of 1920$\times$1080 and bit-rate of 12 Mbps.
    \item \textbf{Data + Video (D+V):} Combination of data and video traffic as described above.
    \item \textbf{Streaming (S):} A live stream video on Youtube is loaded, with a resolution of 1280$\times$720 and bit rate of 7.5 Mbps. 
    \item \textbf{Data + Streaming (D+S):} Combination of data and streaming traffic as described above.
\end{itemize}

We use two Google Pixel 3 phones as LAA and Wi-Fi clients, equipped with NSG and analiti, respectively. In most cases, we observed that all available LAA resource blocks were allocated to our device: this is possibly due to the lack of LAA-capable consumer devices at the time of measurement since LAA is only available in newly released phones which tend to be more expensive.  This assumption that there is no other LAA device allowed us to isolate the LAA/Wi-Fi coexistence experiment to only our devices. Furthermore, we can perform a semi-controlled or emulated Wi-Fi/Wi-Fi coexistence experiment due to the assumed absence of other LAA device. First,  we turn LAA  ON and request different types of traffic. We collect data on SINR, RB allocation, throughput, and the number of Wi-Fi devices on the same channel as LAA. Next, we do the emulated Wi-Fi/Wi-Fi coexistence by turning LAA OFF and repeat the same measurements to measure the performance over Xfinity Wi-Fi without LAA on the same channel. We analyze coexistence performance based on a number of metrics as discussed next.

\subsection{Measurement results and discussions}

\subsubsection{Average SINR and TXOP in the unlicensed band}

Fig.~\ref{asinr} shows the average LAA SINR on Channel 149, 153 and 157. We see that none of the unlicensed channels are enabled by LAA for real-time video streaming since it is more difficult to guarantee the required QoS on the unlicensed channels due to potential interference and the need to implement LBT. We also see that the SINR on Channel 157 (\emph{i.e.,} 2.9 dB) is lower compared to the other channels, for all other traffic types. The reason for this is clear from Fig.~\ref{figuretabledep}b which shows that there are a larger number of Wi-Fi devices on Channel 157 compared to the other unlicensed channels. This increased level of Wi-Fi usage leads to a lower SINR as well as reduced TXOP from 8 to 3 ms, as shown on Fig.~\ref{atxop}.

\subsubsection{Resource Block (RB) allocation and throughput}

Fig.~\ref{arb} shows the relative number of RBs allocated over the licensed channel and unlicensed channels. Since there are 3 unlicensed channels, it is natural that the total number of RBs allocated to unlicensed is greater than that allocated to licensed, but it also indicates that licensed carriers can easily more than triple their downlink throughput as shown in Fig.~\ref{ath}, even up to nine times. This increase in throughput, just by aggregating 3 unlicensed channels, will drive increased LAA deployments in densely populated areas.

\subsubsection{Wi-Fi AP behavior in the presence of LAA}

Fig.~\ref{figuretabledep}b compares the number of connected Wi-Fi clients (as measured by the analiti app) on the corresponding LAA channels when Wi-Fi coexists with Wi-Fi (W/W) compared to when Wi-Fi coexists with LAA (W/L), for different traffic types. The analiti app counts Wi-Fi devices based on the signaling messages present in the beacon messages sent by those APs that support QoS. Since all APs may not do so, this count may be smaller than actual numbers of devices present and hence the result is reported as, for example, ``9+''. For each row, we assume that both the Wi-Fi and LAA clients under test are requesting the same traffic type. We observe that the number of Wi-Fi clients that can be supported in W/W is much higher than in W/L. This is an indication that perhaps the presence of LAA on a particular channel inhibits Wi-Fi devices from operating on the same channel due to interference and reduced access to the medium due to high TXOP values used by LAA.  This conclusion is corroborated by the observation that a larger number of Wi-Fi clients are supported when LAA is streaming video since for that traffic type, LAA is not using the unlicensed channel. We have never observed LAA, irrespective of the operator, dynamically adapting its unlicensed channel usage in response to Wi-Fi usage of the same channels. This is clearly quite different from the manner in which Wi-Fi coexists with Wi-Fi on the same channel.

\subsubsection{Impact of Delay in LAA and Wi-Fi Coexistence}

Fig.~\ref{delay} shows the average delay at the Wi-Fi client observed in Wi-Fi/Wi-Fi coexistence and Wi-Fi/LAA coexistence. We see clearly that there is less delay in Wi-Fi/Wi-Fi coexistence compared to Wi-Fi/LAA coexistence, except for streaming video. In all cases where data is a part of the traffic load (D, D+V and D+S), we see increased delay over Wi-Fi when coexisting with LAA due to the maximum LAA TXOP of 8 ms which results in Wi-Fi getting fewer opportunities to access the channel.

\section{Conclusion}

In this paper, we report results and analyses from the first comprehensive measurement campaign of LAA deployments in a major metropolitan area conducted over several months. We have complemented commercially available tools such as Network Signal Guru and analiti, with a newly developed SigCap app that allows us to quickly capture geotagged cellular and Wi-Fi network data that can be used to visualize RSSI and coverage. The SigCap app and the data collected in the Chicago area are openly available from our website. 

This measurement campaign revealed a number of interesting aspects of LAA that have not been adequately addressed in the research literature to date: (i) all LAA deployments we encountered aggregate 3 unlicensed channels (\emph{i.e.}, multiple LAA narrow-band channel), thus potentially creating a larger impact to Wi-Fi operations (ii) the majority of Wi-Fi deployments today are 40 MHz and 80 MHz, while most analyses have focused on 20 MHz coexistence of Wi-Fi and LAA (iii) the multi-channel Wi-Fi and multi-channel LAA access schemes with different energy detection thresholds create coexistence scenarios which have not been  studied (iv) even though LAA deployments are primarily outdoors and Wi-Fi deployments are indoors, the signal strength of both at client devices operating outdoors is comparable, leading to increased coexistence and hidden node problems and (v) the impact of LTE-LAA on Wi-Fi latency due to the larger TXOPs being used by LAA needs further study and analysis. We continue to make measurements in the Chicago area and our future research will focus on performing coexistence experiments using Wi-Fi APs deployed in a controlled fashion in coverage areas of existing LAA deployments on university campuses in the Chicago area. We hope that the lessons learned from these on-the-ground measurements will enable a deeper understanding of coexistence in present and future unlicensed bands using 802.11ax and 5G NR.

\end{document}